\documentclass[12pt,a4paper]{article}
\usepackage[latin2]{inputenc}
\usepackage{amsmath}
\usepackage{amsfonts}
\usepackage{amssymb}
\usepackage{graphicx}
\usepackage{slashed}
\begin{document}

\title{Particle Dynamics and Lie-algebraic type of Non-commutativity of space-time}
\author{Partha Nandi\footnote{parthanandi@bose.res.in}, Sayan Kumar Pal\footnote{sayankpal@bose.res.in} and Ravikant Verma\footnote{Email:ravikant.verma@bose.res.in, ravikant.uohyd@gmail.com}\\
{\it S. N. Bose National Centre for Basic Sciences,}\\{\it  JD Block, Sector III, Salt Lake, Kolkata 700106, India}}
\maketitle
\begin{abstract}
In this paper, we present the results of our investigation relating particle dynamics and non-commutativity of space-time by using Dirac's constraint analysis. In this study, we re-parameterise the time $t=t(\tau)$ along with $x=x(\tau)$ and treat both as configuration space variables. Here, $\tau$ is a monotonic increasing parameter and the system evolves with this parameter. After constraint analysis, we find the deformed Dirac brackets similar to the $\kappa$-deformed space-time and also, get the deformed Hamilton's equations of motion. Moreover, we study the effect of non-commutativity on the generators of Galilean group and Poincare group and find undeformed form of the algebra. Also, we work on the extended space analysis in the Lagrangian formalism. We find the primary as well as the secondary constraints. Strikingly on calculating the Dirac brackets among the phase space variables, we obtain the classical version of $\kappa$-Minkowski algebra.

\end{abstract}

\section{Introduction}
In recent time, non-commutative physics is a very interesting area of research. Many models of non-commutativity have been proposed such as Moyal space, fuzzy sphere, $\kappa$-Minkowski space-time, Snyder space, etc. Few years back, 't Hooft has viewed quantum gravity as a classical dissipative system\cite{th}. Following his prescription,  spatial non-commutative structure was obtained for planar systems through Dirac brackets\cite{rbs}. In theoretical physics, Dirac bracket is a mathematical tool and it was introduced by Dirac to handle the proper quantisation of the constraint system\cite{dirac} which is the generalisation of the Poisson bracket in classical mechanics for constraint systems. Quantisation of such systems can be proceeded by setting the fundamental commutator bracket equal to the $i\hbar$ times the Dirac bracket. Many developments involving re-parameterisation in quantum mechanics start with the lagrangian formalism of classical mechanics and use Dirac quantisation.

The Planck length plays an important role to understand the physics in the microlevel. The first example of the space-time non-commutativity was introduced by Snyder in 1947\cite{snyder} and it was based on the deformation of the Heisenberg algebra of quantum mechanics. It is believed that non-commutativity of space-time describes the structure of the space-time at the Planck scale and attracted wide attention\cite{nc1,nc2,nc3,nc4,nc5,nc6,p1,p2,p3,p4,p5,p6,p7,p8,p9,p10,p11}. A connection between dissipation and noncommutativity was investigated in\cite{partha}. For canonical non-commutative case, deformed non-relativistic and relativistic symmetries were studied in\cite{RBKK}. The theory of quantum gravity suggest a minimal length scale and the location of the particles in the space-time structure at this scale may become fuzzy\cite{fu}. A rigorous formulation of space-time non-commutative quantum mechanics was done in\cite{PNS}.

In this paper, we did not talk about the microlevel physics. Here, we present all results classically. By using the well known approach of the constrained Hamiltonian system\cite{dirac}, many people have studied the dynamics of the classical particle\cite{n1,n2,n3,syd}. This paper is organised as follows. In the next section, we briefly talk about the $\kappa$-deformed space-time and study the classical dynamics of the free particle and the harmonic oscillator by using the constrained Hamiltonian approach introduced by Dirac. First, we re-parameterise time and treat it as a phase space variable and we fix the gauge condition and find the deformed Dirac bracket and the Hamilton equation of motion. This deformed Dirac bracket is similar to the classical version of $\kappa$-deformed space-time. We present the effect of non-commutativity on the generators of Galilean group in section III. In section IV, we discuss the representation of the Galilean generator from the Noether's prescription of the Lagrangian. In section V, we take up the issues of the relativistic free particle and investigate how does non-commutativity of the space-time arise in this case and also we study the Poincare algebra of the relativistic free particle. Also, we study the Dirac algebra among the phase space variables without fixing the gauge symmetry by the extended phase space Lagrangian analysis for non-relativistic case in section VI and in section VII, we present our concluding remarks.

\section{$\kappa$-deformation of space-time}

In this section, we give a brief summary of the classical nature of the $\kappa$-deformed space-time and we study the classical dynamics of the free particle and the harmonic oscillator viewed as a constrained system. The fundamental Planck length $l_p$ plays an important role in the micro level physics like quantum gravity but in special theory of relativity this fundamental length scale is absent due to the length contraction. This is because in the special theory of relativity, two different detectors do not detect the same length. But if the Planck length is the fundamental length then it should be same in all inertial frame but this does not follow the special theory of relativity. With this idea, Amelino-Camelia developed a modified form of the special theory of relativity which is called the deformed special theory of relativity (DSR)\cite{ac}. This modified form of the special theory of the relativity requires another fundamental constant in addition to the speed of light $c$ and this new fundamental constant is related to the Planck length. $\kappa$-space-time which emerge in the low energy quantum gravity model is one of the example of non-commutative space-time and which is also connected to the deformed special relativity.

From the classical point of veiw, the phase space variables satisfy the Heisenberg algebra with the poisson brackets as

\begin{equation}
\{x^i , x^j\}=0, ~~\{x^i, p^j\}=\delta^{ij},~~~~\{p^i,p^j\}=0,
\end{equation}
where $x^i$ and $p^j$ are the phase space variables. For the $\kappa$-deformed space-time classical Heisenberg algebra gets deformed as\cite{p5}
\begin{equation}
\{x^0 , x^i\}=ax^i, ~~\{x^i, p^j\}=\delta^{ij},~~~~\{p^i,p^j\}=0,
\end{equation}
here $a$ is the deformation parameter. In the limit $a\rightarrow 0$, we get back the commutative Heisenberg algebra.

\subsection{Non-relativistic Free Particle}

In this sub-section, we study the classical dynamics of the free particle by using the constrained Hamiltonian formalism. We start with considering the non-relativistic action for an arbitrary potential as\cite{A.D}
\begin{equation}
S[x(t)]=\int_{t_1}^{t_2} dt L\left(x,\frac{dx}{dt}\right),~~~L=\frac{1}{2}m\left(\frac{dx}{dt}\right)^2 - V(x,t).
\end{equation}
Now we re-parameterise the time $t=t(\tau)$ along with $x=x(\tau)$ and treat as a phase space variables and then above equation can be re-written as
\begin{equation}
S[x(\tau),t(\tau)]=\int_{\tau_1}^{\tau_2} d\tau L_\tau(x,\dot{x},t,\dot{t}),~~~L_\tau(x,\dot{x},t,\dot{t})=\dot{t}L\left(x,\frac{\dot{x}}{\dot{t}}\right),
\end{equation}
where $\dot{t}=\frac{dt}{d\tau}$, $\dot{x}=\frac{dx}{d\tau}$ and $\tau$ is the new evolution parameter. Now the canonical momentum corresponding to phase space variables $t(\tau)$ and $x(\tau)$ with $L_\tau=\frac{1}{2}m\frac{\dot{x}^2}{\dot{t}} - \dot{t} V(x,t)$ is given as
\begin{equation}
p_x =\frac{dL_\tau}{d\dot{x}}=m\frac{\dot{x}}{\dot{t}}=m\left(\frac{dx}{dt}\right)
\end{equation}
and
\begin{equation}
p_t =\frac{dL_\tau}{d\dot{t}}=-\frac{1}{2} m\frac{\dot{x}^2}{\dot{t}^2} - V(x,t)=-\frac{p_x^2}{2m}-V(x,t)=-H , \nonumber
\end{equation}
\begin{equation}
p_t + H =p_t + \frac{p_x^2}{2m} + V(x,t)=0.
\end{equation}
Thus, we have a primary constraint obtain from above as
\begin{equation}
\phi_1=p_t + \frac{p_x^2}{2m} +V(x,t) \approx 0.
\end{equation}
Where $\approx 0$ is the equality in the weak sense\cite{dirac}. As happens for a time re-parameterise theory, the canonical Hamiltonian vanishes:
\begin{equation}
H_c = p_t \dot{t}+p_x \dot{x}-L_\tau =0.\label{H_c}
\end{equation}
All the first class constraints act as the generator of the gauge transformation. Now classically if $\cal{G}$ is the generator of the infinitesimal transformation then for any phase space variable $F$
\begin{equation}
\delta F = \{F, \cal{G}\}_{PB}.
\end{equation}
Also, $F$ is invariant if and only if $\delta F=0$ and here $H_0$ (free particle Hamiltonian) is gauge invariant because $\{H_0, \phi_1\}_{PB}=0$ (for free particle case). Now we take the gauge fixing condition for a free particle as
\begin{equation}
\phi_2 =t-\tau-f(x,p_x) \approx 0.
\end{equation}
Also after gauge fixing, $\phi_1$ and $\phi_2$ are the second class constraints because $\{\phi_1 , \phi_2\}_{PB}\neq 0$. Let us now calculate the $\{\phi_1 , \phi_2\}_{PB}$ as
\begin{eqnarray}
\{\phi_1 , \phi_2\}_{PB} &=& \{p_t +H_0, t-\tau - f(x,p_x)\} \nonumber \\
&=& -1 + \frac{\partial f}{\partial x} \frac{\partial H_0}{\partial p_x}\nonumber \\
&=& -1 + \frac{p_x}{m}\frac{\partial f}{\partial x} =C_{12}.
\end{eqnarray}
Now we define the constraint matrix as
\begin{equation}
C_{\alpha \beta}=\begin{pmatrix}
0 & -1 + \frac{p_x}{m}\frac{\partial f}{\partial x}\\
1 - \frac{p_x}{m}\frac{\partial f}{\partial x} & 0 
\end{pmatrix},
\end{equation}
and
\begin{equation}
(C^{-1})_{\alpha \beta}=\begin{pmatrix}
0 & \frac{1}{1  -\frac{p_x}{m}\frac{\partial f}{\partial x}}\\
\frac{1}{-1 + \frac{p_x}{m}\frac{\partial f}{\partial x}} & 0 
\end{pmatrix}=C^{\alpha\beta}.
\end{equation}
Now we construct the Dirac bracket defined as
\begin{equation}
\{M,N\}_{DB}=\{M,N\}_{PB}-\{M,\phi_\alpha \}_{PB} C^{\alpha\beta} \{\phi_\beta,N\}_{PB},\label{o}
\end{equation}
where $M$ and $N$ are any pair of phase space variables and $C^{\alpha\beta}$ is the inverse matrix. Now by using the gauge fixing condition with the choice of $f=-axp_x$ and after straight forward calculation, we get
\begin{equation}
\{t,x\}_{DB}=\frac{ax}{1+2aE},~~~~~~E=\frac{p_x^2}{2m}.
\end{equation}
Similarly,
\begin{equation}
\{t,p_t\}_{DB}=\frac{2aE}{1+2aE},
\end{equation}
\begin{equation}
\{p_x,t\}_{DB}=\frac{ap_x}{1+2aE},
\end{equation}
and
\begin{equation}
\{x,p_x\}_{DB}=\frac{1}{1+2aE}.
\end{equation}
From above we can get
\begin{equation}
\{t,x\}_{DB}=\lambda x \label{t}
\end{equation}
where constant $\lambda = \frac{a}{1+2aE}$. Above equation is similar to the classical version of the $\kappa$-deformed space-time. In the limit $a\rightarrow 0$, we get back the undeformed results as in the usual classical mechanics.

Now we want to compute the equation of the motion on the constraint surface for the free particle in the $\kappa$-deformed space-time. Since we know from Eqn.(\ref{H_c}) that $H_c\approx 0$ then we can not write the Hamilton equation of motion and because $H_0$ is a gauge invariant quantity(i.e. $\{H_0,\phi_1\}_{PB}=0$). Now after gauge fixing, we have a pair of second class constraints $\phi_1$ and $\phi_2$ and now we look for a new form of the Hamiltonian after gauge fixing as
\begin{equation}
H_0^\prime =H_0 +\lambda_1 \phi_1 +\lambda_2 \phi_2,\label{h1}
\end{equation}
where $\lambda_1$ and $\lambda_2$ are Lagrange's multipliers such that
\begin{equation}
\{H_0^\prime , \phi_1\}_{PB}= 0,~~~\{H_0^\prime , \phi_2\}_{PB}= 0.
\end{equation}
Now
\begin{eqnarray}
\{H_0 +\lambda_1 \phi_1 +\lambda_2 \phi_2 , \phi_1\}_{PB}&=&0\nonumber \\
\{H_0,\phi_1\}_{PB} + \lambda_2\{\phi_2 ,\phi_1\}_{PB} &=& 0,
\end{eqnarray}
since $\{\phi_2 ,\phi_1\}_{PB}\neq 0$ and $\{H_0,\phi_1\}_{PB}=0$ then $\lambda_2 =0$ and now
\begin{eqnarray}
\{H_0 +\lambda_1 \phi_1 +\lambda_2 \phi_2 , \phi_2\}_{PB}&=&0\nonumber \\
\{H_0,\phi_2\}_{PB} + \lambda_1\{\phi_1 ,\phi_2\}_{PB} &=& 0,
\end{eqnarray}
then
\begin{equation}
\lambda_1 = -\frac{\{H_0,\phi_2\}}{\{\phi_1 ,\phi_2\}}=-\frac{ap_x^2}{m\left(1+\frac{ap_x^2}{m}\right)}.
\end{equation}
Now substitute $\lambda_1$ and $\lambda_2$ in Eqn.(\ref{h1}) and we find
\begin{equation}
H_0^\prime = H_0\left( 1- \frac{2a\phi_1}{1+2aE}\right).
\end{equation}
When we are working with the Dirac brackets then we have $H_0^\prime=H_0$ because the Dirac brackets imply a strong imposition of the second class constraints and then the equation of motion on the constraint surface for free particle will be 
\begin{equation}
\dot{x}=\frac{dx}{d\tau}=\frac{\partial x(\tau)}{\partial \tau} +\{x,H_0\}_{DB},~~~~~~\dot{t}=\frac{dt}{d\tau}=\frac{\partial t}{\partial \tau}  + \{t,H_0\}_{DB},
\end{equation}
after the straight forward calculation, we find
\begin{equation}
\dot{x}=\frac{p_x}{m+ap_x^2},~~~\dot{t}=\frac{m}{m+ap_x^2}.
\end{equation}
These are the Hamilton equations of the motion of the free particle on the constraint surface of the system in the deformed space-time and in the limit $a\rightarrow 0$, we get back the usual classical results. Trajectory of free particle on the constraint surface of the sytem is modified. 

\subsection{Harmonic Oscillator}

Here, we study the classical dynamics of the harmonic oscillator as we studied for free particle in the previous sub-section by using constrained Hamiltonian approach. The Lagrangian for the harmonic oscillator in terms of the usual coordinates is given as
\begin{equation}
L=\frac{1}{2m}(p_q^2 - m^2\omega^2 q^2),
\end{equation}
corresponding action is
\begin{equation}
S=\int L(q,\dot{q},t) dt=\int (\dot{q}p_q -H(p_q,q,t))dt.
\end{equation}
Now let us consider the canonical transformation 
\begin{eqnarray}
\dot{q}p_q -H(p_q,q,t)&=&P\dot{Q}-\tilde{H} + \frac{dF(q,Q,t)}{dt},\nonumber \\
\dot{q}p_q -H(p_q,q,t)&=&P\dot{Q}-\tilde{H} +\frac{\partial F}{\partial Q}\frac{dQ}{dt}+\frac{\partial F}{\partial q}\frac{dq}{dt}+\frac{\partial F}{\partial t},
\end{eqnarray}
where $F$ is the generating function of the first kind and from the above equation, we find
\begin{equation}
P=-\frac{\partial F}{\partial Q},~~~~p_q=\frac{\partial F}{\partial q}.
\end{equation}
Also, after transformation from old canonical variables to a new set of canonical variables, action can be re-written as
\begin{eqnarray}
S&=&\int \left(\dot{q}p_q -H(p_q,q,t)-\frac{dF(q,Q,t)}{dt}\right)dt,\nonumber \\
S&=&\int \left(\dot{q}p_q -H(p_q,q,t)-\frac{\partial F}{\partial Q}\frac{dQ}{dt}-\frac{\partial F}{\partial q}\frac{dq}{dt}-\frac{\partial F}{\partial t}\right)dt,\nonumber\\
S&=&\int \left(\dot{Q}P -H(p_q,q,t)-\frac{\partial F}{\partial t}\right)dt,\nonumber\\
S&=&\int (\dot{Q}P -\tilde{H}) dt,
\end{eqnarray}
where $\tilde{H}=H(p_q,q,t)+\frac{\partial F(Q,P)}{\partial t}$(for harmonic oscillator) then $\tilde{H} = H$. Now one can define the Hamiltonian of the harmonic oscillator as
\begin{equation}
H=\frac{p_q^2}{2m}+\frac{1}{2}m\omega^2 q^2,\label{g1}
\end{equation}
here $q$ and $p_q$ in terms of the new set of canonical variables $(Q,P)$ are given as
\begin{eqnarray}
q&=&\sqrt{\frac{2P}{m\omega}} \sin Q\nonumber\\
p_q&=&\sqrt{2m\omega P} \cos Q.
\end{eqnarray}
Using $q$ and $p_q$ in Eqn.(\ref{g1}), we get
\begin{equation}
H=\omega P.
\end{equation}
Now one can re-parameterise time $t=t(\tau)$ as a new phase space variable along with $x(\tau)$ and action can be re-written as
\begin{equation}
S=\int \left(\frac{\dot{Q}}{\dot{t}}P -\tilde{H}\right) \dot{t} d\tau,
\end{equation}
here $\left(\frac{\dot{Q}}{\dot{t}}P -\tilde{H}\right) \dot{t}=L_\tau$ and one can calculate the canonical momentum as
\begin{equation}
p_t=\frac{\partial L_\tau}{\partial \dot{t}}=-\tilde{H},~~P=\frac{\partial L_\tau}{\partial \dot{Q}},
\end{equation}
Also, the primary constraint in this theory is obtained from above as
\begin{equation}
\phi_1 = p_t + \tilde{H} \approx 0.\label{p1}
\end{equation}
Now let us fix the gauge 
\begin{equation}
\phi_2 = t-\tau -f(Q,P)\approx 0,\label{p2}
\end{equation}
then in this case the component of a constraint matrix is given as
\begin{equation}
C^\prime_{12}=\{\phi_1 ,\phi_2\}_{PB}=-1+\frac{\partial H}{\partial P}\frac{\partial f}{\partial Q},
\end{equation} 
also, we can construct the constraint matrix as
\begin{equation}
C^\prime_{\alpha \beta}=
\begin{pmatrix}
0 & -1+\frac{\partial H}{\partial P}\frac{\partial f}{\partial Q}\\
1-\frac{\partial H}{\partial P}\frac{\partial f}{\partial Q}&0
\end{pmatrix}
\end{equation}
and inverse of a constraint matrix is given as
\begin{equation}
(C^{\prime{-1}})_{\alpha\beta}=
\begin{pmatrix}
0& \frac{1}{1-\frac{\partial H}{\partial P}\frac{\partial f}{\partial Q}}\\
\frac{-1}{1-\frac{\partial H}{\partial P}\frac{\partial f}{\partial Q}}&0
\end{pmatrix}=C^{\prime\alpha\beta}.\label{p3}
\end{equation}
Now using Eqns.(\ref{p1}),(\ref{p2}) and (\ref{p3}) with the choice of $f(Q,P)=-aQP$ in Eqn.(\ref{o}) and after straight forward calculation, we get the Dirac brackets between phase space variables as
\begin{equation}
\{t,Q\}_{DB}=\frac{aQ}{1+ aH},~\{t,P\}_{DB}=\frac{-aP}{1+ aH},~\{Q,P\}_{DB}=\frac{1}{1+ aH}.
\end{equation}
Since Hamiltonian $H$ is conserved with respect to time then in this case we can write $H=E$ and above equation re-written as
\begin{equation}
\{t,Q\}_{DB}=\frac{aQ}{1+ aE},~\{t,P\}_{DB}=\frac{-aP}{1+ aE},~\{Q,P\}_{DB}=\frac{1}{1+ aE}.
\end{equation}
These Dirac brackets of the phase space variables look like similar to the $\kappa$-deformed space-time in classical sense. Now since $\tilde{H}$ is a gauge invariant quantity(i.e. $\{\tilde{H},\phi_1\}_{PB}=0$) and after gauge fixing, we have a pair of second class constraints $\phi_1$ and $\phi_2$ and we look for a new form of the Hamiltonian after gauge fixing as
\begin{equation}
H^{\prime} =\tilde{H} +\lambda_1 \phi_1 +\lambda_2 \phi_2,\label{h11}
\end{equation}
where $\lambda_1$ and $\lambda_2$ are Lagrange's multipliers such that
\begin{equation}
\{H^\prime , \phi_1\}_{PB}= 0,~~~\{H^\prime , \phi_2\}_{PB}= 0,
\end{equation}
Now
\begin{eqnarray}
\{\tilde{H} +\lambda_1 \phi_1 +\lambda_2 \phi_2 , \phi_1\}_{PB}&=&0\nonumber \\
\{\tilde{H},\phi_1\}_{PB} + \lambda_2\{\phi_2 ,\phi_1\}_{PB} &=& 0,
\end{eqnarray}
since $\{\phi_2 ,\phi_1\}_{PB}\neq 0$ and $\{\tilde{H},\phi_1\}_{PB}=0$ then $\lambda_2 =0$ and now
\begin{eqnarray}
\{\tilde{H} +\lambda_1 \phi_1 +\lambda_2 \phi_2 , \phi_2\}_{PB}&=&0\nonumber \\
\{\tilde{H},\phi_2\}_{PB} + \lambda_1\{\phi_1 ,\phi_2\}_{PB} &=& 0,
\end{eqnarray}
then
\begin{equation}
\lambda_1 = -\frac{\{\tilde{H},\phi_2\}}{\{\phi_1 ,\phi_2\}}=-\frac{a\omega P}{(1+aH)}.
\end{equation}
Now substitute $\lambda_1$ and $\lambda_2$ in Eqn.(\ref{h11}) and we find
\begin{equation}
H^{\prime} = H\left( 1- \frac{a\phi_1}{1+aE}\right).
\end{equation}
If we are working with the Dirac brackets then we have $H^\prime=H$ because the Dirac brackets imply a strong imposition of the second class constraints and then the equation of motion on the constraint surface will be
\begin{equation}
\dot{Q}=\frac{dQ}{d\tau}=\frac{\partial Q(\tau)}{\partial \tau} +\{Q,H\}_{DB},~~~~~~\dot{t}=\frac{dt}{d\tau}=\frac{\partial t}{\partial \tau}  + \{t,H\}_{DB},
\end{equation}
after the straight forward calculation, we find
\begin{equation}
\dot{Q}=\frac{\omega}{1+a\omega P},~~~\dot{t}=\frac{1}{1+a\omega P}.
\end{equation}
Initially(before time re-parameterisation), $Q=\omega t$, $\dot{Q}=\frac{dQ}{dt}=\omega$ and now after the time re-parameterisation and gauge fixing condition, 
\begin{equation}
Q=\omega t,~~~ \dot{Q}=\frac{dQ}{d\tau}=\omega \frac{dt}{d\tau}=\omega \dot{t}=\frac{\omega}{1+a\omega P}
\end{equation}
and
\begin{equation}
Q=\frac{\omega}{1+a\omega P}\tau=\omega^\prime \tau,
\end{equation}
where $\omega^\prime=\frac{\omega}{1+a\omega P}$. This is modified frequency due to time re-parameterisation and gauge fixing and is less than the frequency $\omega$ in the commutative case.

\section{Effect of Non-commutativity on Representations of Galilean Generators}

In this section, we study the effect of non-commutativity on the generators of Galilean group and algebra. The generators of this group consist of angular momentum $J$, translation $P$ and boost $G$, which have the form in commutative case as\cite{n1}
\begin{equation}
J=\epsilon_{ij}x_ip_j,~~~P_i=p_i,~~~~G_i=mx_i-t p_i,\label{tt}
\end{equation}
where $t$ is the evolution parameter and these generators satisfy the following algebra
\begin{equation}
\{P_i,P_j\}=0,~~~\{J,J\}=0,~~~~~\{G_i,G_j\}=0,
\end{equation}
\begin{equation}
\{P_i,J\}=-\epsilon_{ij}P_j,~~\{G_i,J\}=-\epsilon_{ij}G_j,~~\{P_i,G_j\}=-m\delta_{ij}.
\end{equation}
Since in this paper, we re-parameterise the time $t(\tau)$ and treat as phase space variable in addition to $x(\tau)$ and here $\tau$ is the new evolution parameter. Now the Dirac brackets among the phase space variables
\begin{equation}
\{t,x\}_{DB}=\lambda x,~~~\{p_x,t\}_{DB}=\lambda p_x,~~~~\{x,p_x\}_{DB}=\frac{1}{1+2aE},
\end{equation}
where $t=\tau-axp_x$, $\lambda =\frac{a}{1+2aE}$. Here $t(\tau)$ and $x(\tau)$ are the phase space variables. Now we can re-write the generators in terms of new evolution parameter in (1+1) dimension as
\begin{equation}
J=0,~~~P=p_x,~~~~G=mx-(\tau -axp_x)p_x,\label{kk}
\end{equation}
and satisfy the $\kappa$-undeformed algebra as
\begin{eqnarray}
\{P,P\}_{DB}&=&0,~~\{J,J\}_{DB}=0,~~~\{G,G\}_{DB}=0\\
\{H,G\}_{DB}&=&-P,~~\{P,G\}_{DB}=-m,
\end{eqnarray}
where $H$ is the time translation generator. This algebra is called the undeformed Galilean algebra but the boost generator modified due to the non-commutativity between space and time and remaining generators of the Galilean algebra did not modify due to the deformation of the space-time. In limit $a\rightarrow 0$, we get back the boost generator as in commutative case. Explicit presence of the deformation parameter $a$ in the boost generator implies modification in the infinitesimal symmetry transformation. These are obtain by computing the Dirac bracket of the generator with the phase space variables as
\begin{eqnarray}
\delta t&=&\{t,G\}_{DB}\nonumber\\
&=&\{t,mx-(\tau -axp_x)p_x\}\nonumber\\
&=& m\lambda x+\lambda \tau p_x-a\lambda xp_x^2\nonumber\\ 
\end{eqnarray}
similarly,
\begin{eqnarray}
\delta x&=&\{x,G\}_{DB}~~~~~~~~~~~~~~~~~~~~~~\nonumber\\
&=&\{x,mx-(\tau -axp_x)p_x\}\nonumber\\
&=&-\frac{\tau -2axp_x}{1+2aE}
\end{eqnarray}
and
\begin{eqnarray}
\delta p_x&=&\{p_x,G\}_{DB}~~~~~~~~~~~~~~~~~~~~~~\nonumber\\
&=&\{p_x,mx-(\tau -axp_x)p_x\}\nonumber\\
&=&-m.
\end{eqnarray}
Here, deformation parameter $a$ dependent term shows the modification in the symmetry transformation. Also, these expressions have the correct commutative limit and there is no modification in the symmetry transformation of translation generator $(P)$ because translation generator does not have any modification due to deformation of the space-time.

\section{Noether's Theorem and Boost Generator}

In this section, we present the representation of the Galilean generator from Noether's prescription\cite{n1,n2,n3}. As in this paper, we re-parameterise the time $t(\tau)$ and treat it as a phase space variable. Also, as happens for a time re-parameterised theory, the canonical Hamiltonian is zero. From Eqn.(\ref{H_c}), we can write Lagrangian as
\begin{equation}
L_\tau = \dot{x}p_x + \dot{t}p_t,
\end{equation}
where $\dot{x}=\frac{\partial x}{\partial \tau}$ and $\dot{t}=\frac{\partial t}{\partial \tau}$ and the canonical momenta conjugate to $x$ and $t$ are
\begin{equation}
\pi^x = \frac{\partial L_\tau}{\partial \dot{x}}=p_x,~~~~~\pi^t = \frac{\partial L_\tau}{\partial \dot{t}}=p_t=-\frac{p_x^2}{2m}.
\end{equation}
Now one can write the Generator $G$ from the usual Noether's theorem prescription as
\begin{equation}
G=\pi^x \delta x +\pi^t \delta t-B,~~~\delta L_\tau =\frac{dB}{d\tau},~~~\delta x^\mu =\{x^\mu ,G\},\label{ll}
\end{equation}
now
\begin{eqnarray}
\delta L_\tau &=& \delta\dot{x} p_x + \dot{x}\delta p_x + \delta\dot{t} p_t + \dot{t}\delta p_t\nonumber\\
&=&\frac{d}{d\tau}(\delta x p_x + x\delta p_x + \delta t p_t + t\delta p_t),
\end{eqnarray}
where $B=\delta x p_x + x\delta p_x + \delta t p_t + t\delta p_t$. Now the generator $G$ in Eqn.(\ref{ll}) can be r-ewritten as
\begin{eqnarray}
G&=&p_x \delta x +p_t \delta t-(\delta x p_x + x\delta p_x + \delta t p_t + t\delta p_t)\nonumber\\
&=&-\left( x\delta p_x -t\frac{p_x}{m}\delta p_x\right)= \bigg(mx(\tau)-t(\tau)p_x\bigg) \epsilon = \cal{G}\epsilon,
\end{eqnarray}
where $\cal{G}$ is the Galilean boost generator and given by
\begin{equation}
{\cal{G}}=mx(\tau)-t(\tau)p_x.
\end{equation}
This generator $\cal{G}$ is gauge invariant because $\delta {\cal{G}}=\{{\cal{G}},\phi_1\}=0$. If we fix the gauge condition then $t=\tau - axp_x$(see section II) and the generator $\cal{G}$ will take the form as
\begin{equation}
{\cal{G}}=mx-\tau p_x + axp_x^2,
\end{equation}
here $\tau$ is the new evolution parameter after re-parameterisation of time. This generator has the same form as given in Eqn.(\ref{kk}) and in the limit $a\rightarrow 0$, we get back the generator $\cal{G}$ in commutative case as gevin in Eqn.(\ref{tt}). As we saw in this present section and also, in previous section that boost generator is modified due to choice of the gauge fixing condition and in ours choice of the gauge condition Galilean algebra has the same form as in commutative case, known as undeformed Galilean algebra in deformed space-time as we seen in previous section.

\section{Relativistic Free Particle}

In the earlier sections, we studied the problem of non-relativistic free particle but here in this section, we take up the issues of the relativistic free particle\cite{syd} and find out how does non-commutativity of space-time arise here and also we study the Poincare algebra of the relativistic free particle. The action for the relativistic free particle is written as
\begin{equation}
S_{rel} = -m \int d\tau \sqrt{-\dot{x}^\mu \dot{x}_\mu},
\end{equation}
where $\dot{x}^\mu=\frac{\partial x^\mu}{\partial\tau}$, the Minkowski metrix $\eta_{\mu\nu}=$diag(-1,+1) and $\mu, \nu = 0,1$. As we know, this relativistic action leads to the mass shell constraint or Einstein constraint given by
\begin{equation}
\phi_1 = p^\mu p_\mu +m^2 \approx 0.
\end{equation}
Where $p_\mu$ is canonically conjugate to $x_\mu$ in the Hamiltonian formalism. Here note that $x_\mu$ (including $x^0=t)$ is manifestly in the re-parameterised form and contained in the configuration space. Now we take the gauge fixing condition as
\begin{equation}
\phi_2 = t-\tau -f(x,p_x)\approx 0.
\end{equation}
Then we calculate
\begin{eqnarray}
\{\phi_1,\phi_2\}&=&\{-p_t^2+p_x^2 +m^2,t-\tau-f\}\nonumber\\
&=&-2p_t + 2p_x\frac{\partial f}{\partial x}= C_{12}.
\end{eqnarray}
Now the constraint matrix can be defined for the relativistic free particle as 
\begin{equation}
C_{\alpha \beta}=\begin{pmatrix}
0 & -2p_t + 2p_x\frac{\partial f}{\partial x}\\
2p_t - 2p_x\frac{\partial f}{\partial x} & 0 
\end{pmatrix},
\end{equation}
and
\begin{equation}
(C^{-1})_{\alpha \beta}=\begin{pmatrix}
0 & \frac{1}{2p_t - 2p_x\frac{\partial f}{\partial x}}\\
\frac{1}{-2p_t + 2p_x\frac{\partial f}{\partial x}} & 0 
\end{pmatrix}=C^{\alpha\beta}.
\end{equation}
Now by using Eqn.(\ref{o}) and with the choice of $f=-axp_x$, we get
\begin{equation}
\{t,x\}_{DB}=\frac{ax}{1+aE^\prime},
\end{equation}
Similarly,
\begin{equation}
\{t,p_t\}_{DB}=\frac{-aE^\prime}{1+aE^\prime},
\end{equation}
\begin{equation}
\{p_x,t\}_{DB}=\frac{ap_x}{1+aE^\prime},
\end{equation}
and
\begin{equation}
\{x,p_x\}_{DB}=\frac{1}{1+aE^\prime}.
\end{equation}
Where $E^\prime=\frac{E^2 -m^2}{E}$ and we have used $\{x^\mu,p^\nu\}=\eta^{\mu\nu}$ and also, in this section $p^0=p_t$. For this calculation, we have set $c=1$. On explicitly retaining $c$ till the end of the calculations and then taking the limit $c\rightarrow \infty$, we get back the result for nonrelativistic free particle (see section 2.1 in this paper).

Since in this paper, we are working in (1+1) dimensions, then there is no rotation generator. The translation generators corresponding to time and position and Lorentz boost will be given as
\begin{equation}
P_x=p_x,~~~P_t =p_t,~~~M^{tx}=tp_x-xp_t
\end{equation}
This boost generator is gauge invariant because $\{M^{tx},\phi_1\}=0$. Since here we re-parameterise the time $t$ and treat it as a phase space variable, then Lorentz boost generator in terms of the new evolution parameter $\tau$ can be re-written as
\begin{equation}
M^{tx}=(\tau -axp_x)p_x-xp_t.
\end{equation}
These generators satisfy the following algebra
\begin{equation}
\{P_x,M^{tx}\}_{DB}=P_t,~~~~~\{P_t,M^{tx}\}_{DB}=P_x.
\end{equation}
This algebra is called the undeformed Poincare algebra but only the boost generator gets modified due to the deformation of the space-time. In limit $a\rightarrow 0$, we get back the Lorentz boost generator as in commutative case.

\section{Gauge independent analysis}

In this section, we analyse the Dirac bracket algebra among the dynamical variables without gauge fixing. People have started with a first order deformed action for free particle and studied the Dirac algebra among the configuration space variables which is basically the classical version of the Snyder algebra\cite{n3} and Moyal algebra\cite{AD}. In the earlier sections of this paper, we have obtained an algebra which is similar to the kappa algebra by the use of a proper gauge fixing condition but here for studying this algebra we are not fixing the gauge. Now for any arbitrary potential, we propose a deformed non-relativistic action as
\begin{equation}
S = \int  d\tau \Bigg[ \dot{x}p_x + \dot{t}p_t + \frac{a}{2} \dot{t}p_x^2 + ax \dot{p}_x p_t + a \dot{x}p_x p_t - e (p^2_x + 2mp_t + 2mV(x,t)) \Bigg]\label{zx}
\end{equation}
where $a$ is the deformation parameter and $e(\tau)$ is a Lagrange's multiplier. This is a first order action written in an extended phase space. In the limit $a\rightarrow 0$, we get the Lagrangian given in\cite{RPM} for an arbitrary potential. The canonical conjugate momenta are 
\begin{eqnarray}
\Pi_t &=& \frac{\partial L}{\partial \dot{t}} = p_t + \frac{a}{2}p^2_x \nonumber\\ 
\Pi_x &=& \frac{\partial L}{\partial \dot{x}} = p_x + a p_x p_t \nonumber\\
\Pi_{p_t} &=& \frac{\partial L}{\partial \dot{p_t}} = 0 \nonumber\\
\Pi_{p_x} &=& \frac{\partial L}{\partial \dot{p_x}} = axp_t \nonumber\\
\Pi_e &=& \frac{\partial L}{\partial \dot{e}} = 0 . \nonumber
\end{eqnarray}
These momenta do not involve only the corresponding velocities and therefore all these are to be interpreted as primary constraints\cite{dirac} which are given by
\begin{eqnarray}
\Phi_{1} &=& \Pi_t - p_t - \frac{a}{2}p^2_x \approx 0 \\
\Phi_{2} &=& \Pi_x - p_x - a p_x p_t \approx 0\\
\Phi_{3} &=& \Pi_{p_t} \approx 0\\
\Phi_{4}&=& \Pi_{p_x} -axp_t \approx 0\\
\Phi &=& \Pi_e \approx 0.
\end{eqnarray}
We can re-write the above constraints as
\begin{eqnarray}
\Phi_{0,i} &=& \Pi_i^q -p_{x_i} - \frac{a}{2}\delta_{i,0}~ p^2_{x_1} - a\delta_{i,1}~ p_{x_1} p_{x_0} \approx 0\nonumber\\
\Phi_{1,i} &=& \Pi_i^p -a\delta_{i,1}~ x_1 p_{x_0} \approx 0, \nonumber
\end{eqnarray}
where $x_0 =t,~x_1=x$ and $i$ run from $0$ to $1$. From Eqn.(\ref{zx}), the canonical Hamiltonian can be easily written as
\begin{equation}
H_c = e (p^2_x + 2mp_t + 2mV(x,t))
\end{equation} 
and the total Hamiltonian is given by
\begin{equation}
H_t = e (p^2_x + 2mp_t + 2mV(x,t)) + \gamma \Phi + \gamma_{i,1} \Phi_{i,1} + \gamma_{p_i,2}\Phi_{p_i,2}
\end{equation}
where $\gamma$'s are the Lagrange's multipliers. Here we have only the time consistency of the constraint $\Pi_e \approx  0$ leading to a secondary constraint given by
\begin{eqnarray}
\Psi &=& \{H_t, \Pi_e\} = (p^2_x + 2mp_t + 2mV(x,t))\approx 0\\
\end{eqnarray}
Now we compute the constraint matrix as
\begin{eqnarray}
\Lambda_{ij}^{mn}&=&\begin{pmatrix}
\{\Phi_{0,i},\Phi_{0,j}\} & \{\Phi_{0,i},\Phi_{1,j}\}\\
\{\Phi_{1,i},\Phi_{0,j}\} & \{\Phi_{1,i},\Phi_{1,j}\}
\end{pmatrix}\nonumber\\
&=& \begin{pmatrix}
0 & -\delta_{ij} - a(\delta_{i,0}\delta_{j,1}+\delta_{i,1}\delta_{j,0})p_x \\
\delta_{ij} + a(\delta_{i,0}\delta_{j,1}+\delta_{i,1}\delta_{j,0})p_x & a\epsilon_{ij}x
\end{pmatrix}.
\end{eqnarray} 
Inverse of the constraint matrix $\Lambda_{ij}$ is computed such that $\Lambda_{ik}\Lambda_{kj}^{-1}=\delta_{ij}$ and is given by
\begin{equation}
(\Lambda^{-1})_{ij}^{mn} = \begin{pmatrix}
\frac{a\epsilon_{ij}x}{1-a^2p_x^2} & \frac{\delta_{ij} - a(\delta_{i,0}\delta_{j,1}+\delta_{i,1}\delta_{j,0})p_x}{1-a^2p_x^2} \\
\frac{-\delta_{ij} + a(\delta_{i,0}\delta_{j,1}+\delta_{i,1}\delta_{j,0})p_x}{1-a^2p_x^2} & 0
\end{pmatrix}.
\end{equation} 
From this point, we can compute the various Dirac brackets using the following definition 
\begin{equation}
\{A,B\}_{DB} = \{A,B\}_{PB} - \{A,\Phi_{m,i}\}_{PB}(\Lambda^{-1})_{ij}^{mn} \{\Phi_{n,j},B\}_{PB}.
\end{equation} 
The Dirac brackets among the phase space variables are
\begin{eqnarray}
\{t,x\}_{DB} &=& \frac{ax}{1-a^2p_x^2}\\
\{p_x,t\}_{DB} &=& \frac{ap_x}{1-a^2p_x^2}\\
\{x,,p_x\}_{DB} &=& \frac{1}{1-a^2p_x^2}.
\end{eqnarray}
The obtained Dirac brackets are compatible with the Jacobi identity. Now,if the deformation parameter $a$ is very small, retaining only upto first order in deformation parameter $a$ to get the Dirac brackets as
 \begin{eqnarray}
\{t,x\}_{DB} &=& ax\\
\{p_x,t\}_{DB} &=& ap_x\\
\{x,,p_x\}_{DB} &=& 1.
\end{eqnarray}
This algebra is the classical version of the kappa deformed algebra and also satisfies the Jacobi identity. Here we can also study infinitesimal space-time symmetry transformation using Noether's theorem and get the deformed version of Galilean boost generator as in the previous section. Finally we would like to comment that the system has two first class constraints which indicate that it has a gauge degree of freedom. The generator of the gauge transformation can be written as
\begin{equation}
G=a_1 \Pi_e +a_2 \Psi,
\end{equation}
here, $a_1$ and $a_2$ depend on the evolution parameter $\tau$.

\section{Conclusion}

In this paper, we studied the classical dynamics of the free particle and harmonic oscillator. For studying this, we used the constrained Hamiltonian approach. First, we studied the dynamics of the free particle. For this, first we re-parameterised the time and treated as configuration space variable and then fix the gauge condition and calculated the Dirac brackets between the phase space variables and also presented the modified Hamilton equation of motion due to the deformed Dirac brackets. Secondly, we studied the classical dynamics of the Harmonic oscillator by using the same approach as we used for the free particle. Also, we found the representation of the generators of the Galilean group and their algebra in the deformed space-time and deformed symmetry transformation. We studied the Poincare algebra of the relativistic particle and the form of the algebra is undeformed but Lorentz boost generator is modified due to deformation of the space-time. Moreover, we proposed a deformed action and studied the Dirac algebra among the phase space variables which is similar to the kappa algebra in the classical level up to the first order in the deformation parameter $a$. All the results in this paper have the smooth commutative limit. 

As we have studied the deformed algebra among the phase space variables in the extended phase space by using the deformed action (see Eqn.(\ref{zx})). This will have the implications in the path-integral formalism. Using the deformed action, we are studying the propagator for arbitrary potential and plan to take the harmonic oscillator example. Work along these lines is in progress and shall be reported separately.

\section*{Acknowledgements:}
We would like to thank Prof. Biswajit Chakraborty for suggestions and discussions. S. K. Pal would like to thank UGC India for financial support. Also, we would like to thank to referees for their useful comments
to make this manuscript more effective.\\~\\

\end{document}